# A General Systems Theory for Atmospheric Flows and Atmospheric Aerosol Size Distribution

A. M. Selvam[1]

*Deputy Director (Retired)*

*Indian Institute of Tropical Meteorology, Pune 411 008, India*
Email: amselvam@gmail.com
Web sites: http://amselvam.webs.com
http://amselvam.tripod.com/index.html

**Abstract**

Atmospheric flows exhibit selfsimilar fractal spacetime fluctuations manifested as the fractal geometry to global cloud cover pattern and inverse power law form for power spectra of meteorological parameters such as windspeed, temperature, rainfall etc. Inverse power law form for power spectra indicate long-range spacetime correlations or non-local connections and is a signature of selforganised criticality generic to dynamical systems in nature such as river flows, population dynamics, heart beat patterns etc. The author has developed a general systems theory which predicts the observed selforganised criticality as a signature of quantumlike chaos in dynamical systems. The model predictions are (i) The fractal fluctuations can be resolved into an overall logarithmic spiral trajectory with the quasiperiodic Penrose tiling pattern for the internal structure. (ii) The probability distribution represents the power (variance) spectrum for fractal fluctuations and follows universal inverse power law form incorporating the *golden mean*. Such a result that the additive amplitudes of eddies when squared represent probability distribution is observed in the subatomic dynamics of quantum systems such as the electron or photon. Therefore the irregular or unpredictable fractal fluctuations exhibit quantumlike chaos. (iii) Atmospheric aerosols are held in suspension by the vertical velocity distribution (spectrum). The atmospheric aerosol size spectrum is derived in terms of the universal inverse power law characterizing atmospheric eddy energy spectrum. Model predicted spectrum is in agreement with the following two experimentally determined atmospheric aerosol data sets, (i) SAFARI 2000 CV-580 Aerosol Data, Dry Season 2000 (CARG) (ii) World Data Centre Aerosols data sets for the three stations Ny Ålesund, Pallas and Hohenpeissenberg.

*Key words and phrases*: universal atmospheric aerosol size spectrum, SAFARI 2000 aerosol size spectra, World data center aerosol size spectra, fractal fluctuations in atmospheric flows, chaos and nonlinear dynamics

## 1. Introduction

Information on the size distribution of atmospheric aerosols is important for the understanding of the physical processes relating to the studies in weather, climate, atmospheric electricity, air pollution and aerosol physics. Aerosols affect the radiative balance of the Earth/atmosphere system via the direct effect whereby they scatter and absorb solar and terrestrial radiation, and via the indirect effect whereby they modify the microphysical properties of clouds thereby affecting the radiative properties and lifetime of clouds [15]. At present empirical models for the size distribution of atmospheric suspended

---

[1] Present postal address: Dr.Mrs.A.M.Selvam, B-1 Aradhana, 42/2A Shivajinagar, Pune 411005, India

particulates is used for quantitative estimation of earth-atmosphere radiation budget related to climate warming/cooling trends. The empirical models for different locations at different atmospheric conditions, however, exhibit similarity in shape implying a common universal physical mechanism governing the organization of the shape of the size spectrum. The pioneering studies during the last three decades by Lovejoy and his group [27, 28] show that the particulates are held in suspension in turbulent atmospheric flows which exhibit selfsimilar fractal fluctuations on all scales ranging from turbulence (mm-sec) to climate (kms-years). Atmospheric flows exhibit selfsimilar fractal fluctuations generic to dynamical systems in nature. A general systems theory for the observed fractal space-time fluctuations of dynamical systems [32 - 35] helps formulate a simple model to explain the observed vertical distribution of number concentration and size spectra of atmospheric aerosols. The atmospheric aerosol size spectrum is derived in terms of the universal inverse power law characterizing atmospheric eddy energy spectrum. A universal (scale independent) spectrum is derived for suspended atmospheric particulate size distribution expressed as a function of the golden mean $\tau$ ($\approx$ 1.618), the total number concentration and the mean volume radius (or diameter) of the particulate size spectrum. Knowledge of the mean volume radius and total number concentration is sufficient to compute the total particulate size spectrum at any location. The physical basis and the theory relating to the model are discussed. The model predictions are (i) Fractal fluctuations can be resolved into an overall logarithmic spiral trajectory with the quasiperiodic Penrose tiling pattern for the internal structure. (ii) The probability distribution of fractal space-time fluctuations represents the power (variance) spectrum for fractal fluctuations and follows universal inverse power law form incorporating the *golden mean*. Such a result that the additive amplitudes of eddies when squared represent probability distribution is observed in the subatomic dynamics of quantum systems such as the electron or photon. Therefore the irregular or unpredictable fractal fluctuations exhibit quantumlike chaos. (iii) Atmospheric aerosols are held in suspension by the vertical velocity distribution (spectrum). The normalised atmospheric aerosol size spectrum is derived in terms of the universal inverse power law characterizing atmospheric eddy energy spectrum. Model predicted spectrum is in agreement with the following two experimentally determined atmospheric aerosol data sets, (i) SAFARI 2000 CV-580 Aerosol Data, Dry Season 2000 (CARG) (ii) World Data Centre Aerosols data sets for the three stations Ny Ålesund, Pallas and Hohenpeissenberg. The paper is organized as follows. The current state of knowledge of the size distribution of atmospheric suspended particulates is given in Sec. 2. Sec. 3 contains a brief summary of the observed characteristics of selfsimilar fractal fluctuations in atmospheric flows. Sec. 4 summarizes the general systems theory for fractal space-time fluctuations in atmospheric flows. The normalized (scale independent) atmospheric aerosol size spectrum is derived in Sec. 5. The close correspondence between the basic physical concepts of general systems theory and classical statistical physics are discussed in Sec. 6. Secs. 7 and 8 give respectively, details of observational data sets used for validating the theoretical predictions, and results of analyses of the data sets. The conclusions of the study are given in Sec. 9.

## 2. Atmospheric aerosol size distribution: current state of knowledge

As aerosol size is one of the most important parameters in describing aerosol properties and their interaction with the atmosphere, its determination and use is of fundamental importance. Aerosol size covers several decades in diameter and hence a variety of instruments are required for its determination. This necessitates several definitions of the diameter, the most common being the geometric diameter *d*. The size fraction with $d > 1$-$2$ μm is usually referred to as the coarse mode, and the fraction $d < 1$-$2$ μm is the fine mode. The latter mode can be further divided into the accumulation $d \sim 0.1$-$1$ μm, Aitken $d \sim 0.01$-$0.1$ um, and nucleation $d < 0.01$ μm modes [16].

Husar [18] has summarized the history of aerosol science as follows. The modern science of atmospheric aerosols began with the pioneering work of Christian Junge who performed the first comprehensive measurements of the size distribution and chemical composition of atmospheric aerosols [20 - 23]. Based on tedious and careful size distribution measurements performed over many different parts of the world, Junge and co-workers have observed that there is a remarkable similarity in the gathered size distributions: they follow a power law function over a wide range from 0.1 to over 20 μm in particle radius.

$$\frac{dN}{d\log r} = cr^{-\alpha}$$

The inverse power law exponent α of the number distribution function ranged between 3 and 5 with a typical value of 4. This power-law form of the size distribution became known as the 'Junge distribution' of atmospheric aerosols. In the 1960s the physical mechanisms that were responsible for producing these similarities in the atmospheric aerosol size spectra were not known, although it was clear that homogeneous and heterogeneous nucleation, coagulation, sedimentation and other removal processes were all influential mechanisms. In particular, it was unclear which combination of these mechanisms is responsible for maintaining the observed 'quasi-stationary size distribution' of the size spectra.

Whitby [44] introduced the concept of the multimodal nature of atmospheric aerosol and Jaenicke *et al* [19] added the mathematical formalism used today. Around 1970 - 71, Whitby *et al*. [43] collected and analyzed several size distribution data sets arising from different locations, times, and sampling methods and the broad range of data provided strong evidence that bimodal distribution occurs as a ubiquitous feature of atmospheric aerosols in general, though the causal processes and mechanisms were unclear. Semi-quantitative explanation of the observed fine particle dynamics provided the scientific support for the bimodal concept and became the basis of regional dynamically coupled gas-aerosol models. As pointed out by Whitby [45] and Junge [23] an actual size distribution comes from the sum of single modes. There is an equivalency between the optical properties of a combination of several modes and a representative single mode. From previous work it can reasonably be assumed that aerosol size distributions follow a lognormal distribution [41]. Physical size distributions can be characterized well by a trimodal model consisting of three additive lognormal distributions [46]. Typically, the planetary boundary layer (PBL) aerosol is combination of three modes corresponding to Aitken nuclei, accumulation mode aerosols, and coarse aerosols, the shape of which is often modeled as the sum of lognormal modes [46, 8].

## 3. Selfsimilar fractal fluctuations in atmospheric flows

Atmospheric flows exhibit self-similar fractal fluctuations generic to dynamical systems in nature. Self-similarity implies long-range space-time correlations identified as self-organized criticality [1]. The physics of self-organized criticality ubiquitous to dynamical systems in nature and in finite precision computer realizations of non-linear numerical models of dynamical systems is not yet identified. During the past three decades, Lovejoy and his group [28] have done extensive observational and theoretical studies of fractal nature of atmospheric flows and emphasize the urgent need to formulate and incorporate quantitative theoretical concepts of fractals in mainstream classical meteorological theory. The empirical analyses summarized by Lovejoy and Schertzer [28] directly demonstrate the strong scale dependencies of many atmospheric fields, showing that they depend in a power law manner on the space–time scales over which they are measured. In spite of intense efforts over more than 50 years, analytic approaches have been surprisingly ineffective at deducing the statistical properties of turbulence. Atmospheric science labors under the misapprehension that its basic science issues have long been settled and that its task is limited to the application of known laws — albeit helped by ever larger quantities of data themselves processed in evermore powerful computers and exploiting ever more sophisticated algorithms. Conclusions about anthropogenic influences on the atmosphere can only be drawn with respect to the null hypothesis, i.e. one requires a theory of the natural variability, including knowledge of the probabilities of the extremes at various resolutions. At present, the null hypotheses are classical so that they assume there are no long-range statistical dependencies and that the probabilities are thin-tailed (i.e., exponential). However observations show that cascades involve long-range dependencies and (typically) have fat tailed (algebraic) distributions in which extreme events occur much more frequently and can persist for much longer than classical theory would allow [28].

A general systems theory for the observed fractal space-time fluctuations of dynamical systems developed by the author [32 - 35] helps formulate a simple algebraic model to explain the observed vertical distribution of number concentration and size spectra of atmospheric aerosols. The atmospheric aerosol size spectrum is derived in terms of the universal inverse power law characterizing atmospheric eddy energy spectrum. The physical basis and the theory relating to the model are discussed in Sec. 4. The model predictions are (i) The fractal fluctuations can be resolved into an overall logarithmic spiral trajectory with the quasiperiodic Penrose tiling pattern for the internal structure. (ii) The probability distribution of fractal space-time fluctuations also represents the power (variance) spectrum for fractal fluctuations and is quantified as universal inverse power law incorporating the *golden mean*. Such a result that the additive amplitudes of eddies when squared represent probability distribution is observed in the subatomic dynamics of quantum systems such as the electron or photon. Therefore the irregular or unpredictable fractal fluctuations exhibit quantum-like chaos. (iii) Atmospheric aerosols are held in suspension by the vertical velocity fluctuation distribution (spectrum). The normalized (scale independent) atmospheric aerosol size spectrum is derived in terms of the universal inverse power law characterizing atmospheric eddy energy spectrum. Model predicted spectrum is in agreement with experimentally determined data sets (Sec. 7 and Sec. 8).

## 4. General systems theory for fractal space-time fluctuations in atmospheric flows

The non-deterministic model described below incorporates the physics of the growth of macro-scale coherent structures from microscopic domain fluctuations in atmospheric flows [32 - 35]. In summary, the mean flow at the planetary ABL possesses an inherent upward momentum flux of frictional origin at the planetary surface. This turbulence-scale upward momentum flux is progressively amplified by the exponential decrease of the atmospheric density with height coupled with the buoyant energy supply by micro-scale fractional condensation on hygroscopic nuclei, even in an unsaturated environment [29]. The mean large-scale upward momentum flux generates helical vortex-roll (or large eddy) circulations in the planetary atmospheric boundary layer and is manifested as cloud rows and (or) streets, and meso-scale cloud clusters (MCC) in the global cloud cover pattern. A conceptual model [32 - 35] of large and turbulent eddies in the planetary ABL is shown in Figs. 1 and 2. The mean airflow at the planetary surface carries the signature of the fine scale features of the planetary surface topography as turbulent fluctuations with a net upward momentum flux. This persistent upward momentum flux of surface frictional origin generates large-eddy (or vortex-roll) circulations, which carry upward the turbulent eddies as internal circulations. Progressive upward growth of a large eddy occurs because of buoyant energy generation in turbulent fluctuations as a result of the latent heat of condensation of atmospheric water vapour on suspended hygroscopic nuclei such as common salt particles. The latent heat of condensation generated by the turbulent eddies forms a distinct warm envelope or a micro-scale capping inversion layer at the crest of the large-eddy circulations as shown in Fig. 1.

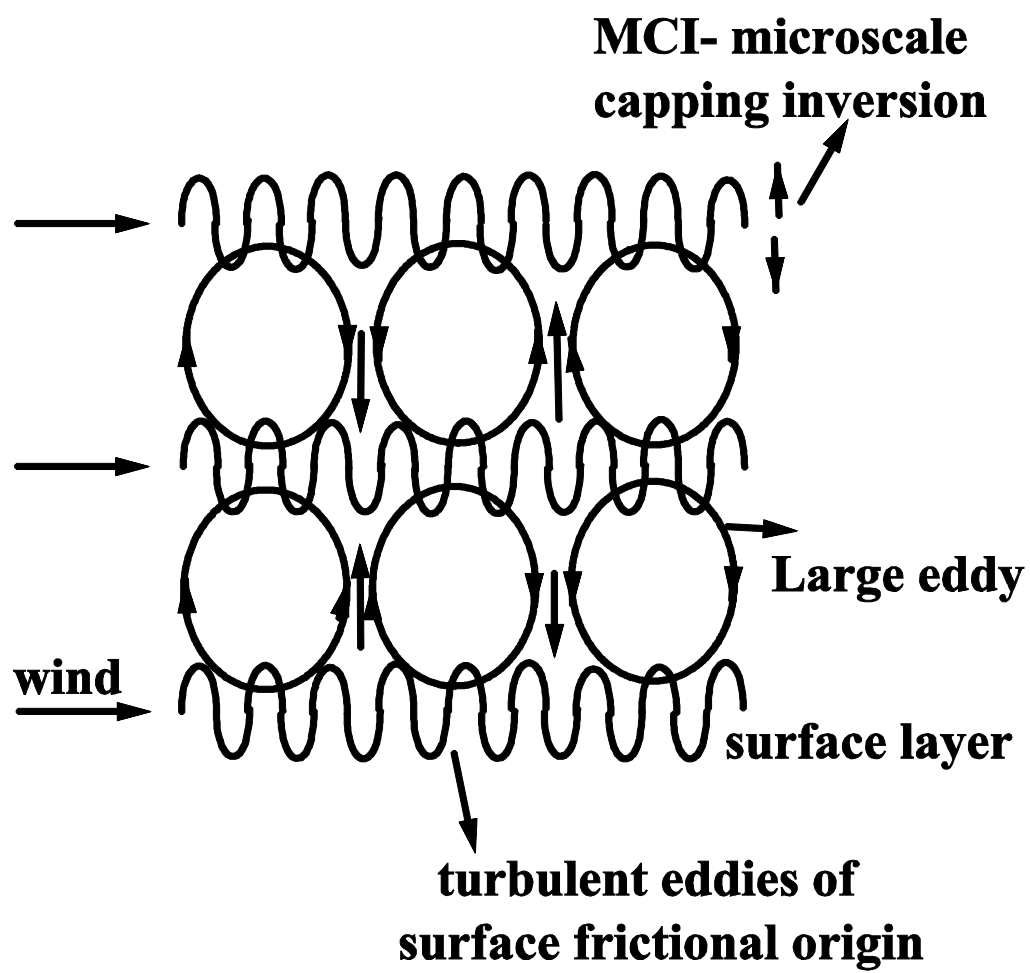

Fig. 1. Micro-scale capping inversion (MCI) layer at the crest of the large-eddy circulations

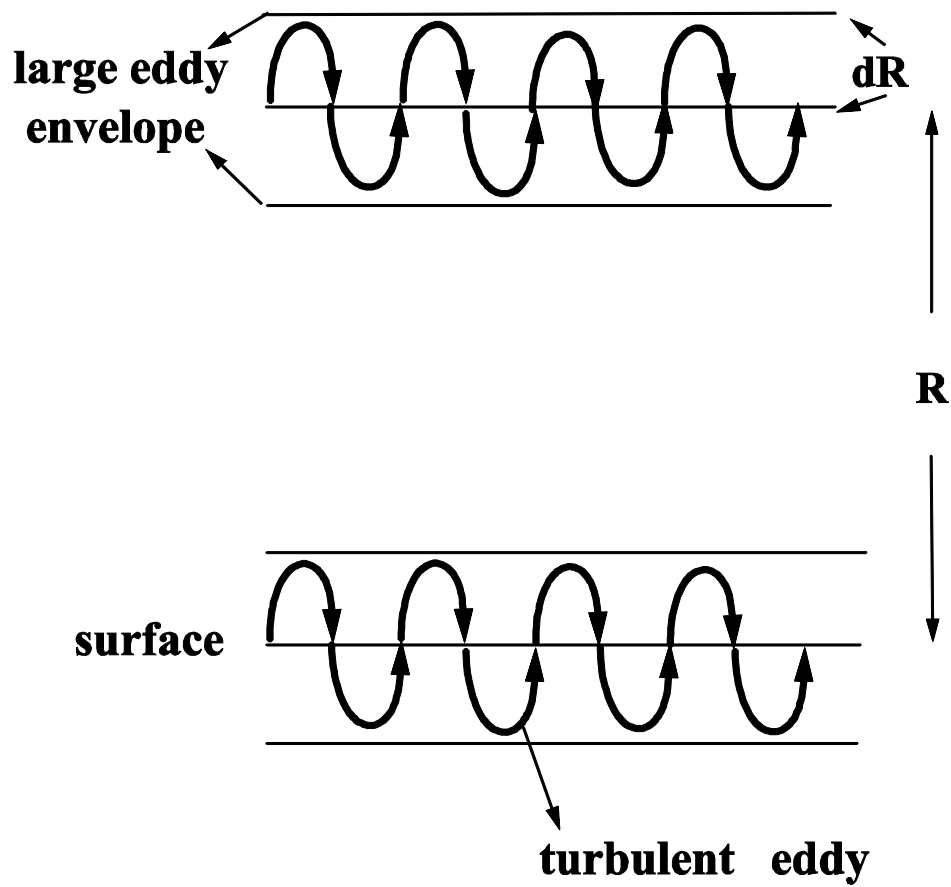

Fig. 2. Progressive upward growth of the large eddy from the turbulence scale at the planetary surface

Progressive upward growth of the large eddy occurs from the turbulence scale at the planetary surface to a height *R* and is seen as the rising inversion of the daytime atmospheric boundary layer (Fig. 2). The turbulent fluctuations at the crest of the growing large-eddy mix overlying environmental air into the large-eddy volume, i.e. there is a two-stream flow of warm air upward and cold air downward analogous to superfluid turbulence in liquid helium [11]. The convective growth of a large eddy in the atmospheric boundary layer therefore occurs by vigorous counter flow of air in turbulent fluctuations, which releases stored buoyant energy in the medium of propagation, e.g., latent heat of condensation of atmospheric water vapour. Such a picture of atmospheric convection is different from the traditional concept of atmospheric eddy growth by diffusion, i.e. analogous to the molecular level momentum transfer by collision [32 - 35].

The generation of turbulent buoyant energy by the micro-scale fractional condensation is maximum at the crest of the large eddies and results in the warming of the large-eddy volume. The turbulent eddies at the crest of the large eddies are identifiable by a micro-scale capping inversion that rises upward with the convective growth of the large eddy during the course of the day. This is seen as the rising inversion of the daytime planetary boundary layer in echosonde and radiosonde records and has been identified as the entrainment zone [3] where mixing with the environment occurs.

In summary, a gravity wave feedback mechanism for the vertical mass exchange between the troposphere and the stratosphere is proposed. The vertical mass exchange takes place through a chain of eddy systems. The atmospheric boundary layer (ABL) contains large eddies (vortex rolls) which carry on their envelopes turbulent eddies of surface frictional origin [32 - 35]. The buoyant energy production by *microscale-fractional-condensation* (MFC) in turbulent eddies is responsible for the sustenance and growth of large eddies [32 - 35, 37].

The buoyant energy production of turbulent eddies by the MFC process is maximum at the crest of the large eddies and results in the warming of the large eddy volume. The turbulent eddies at the crest of the large eddies are identifiable by a *microscale-capping-inversion* (MCI) layer which rises upwards with the convective growth of the large eddy in the course of the day. The MCI layer is a region of enhanced aerosol concentrations. As the parcel of air corresponding to the large eddy rises in the stable environment of the MCI, *Brunt Vaisala* oscillations are generated [32 - 35, 38]. The growth of the large eddy is associated with generation of a continuous spectrum of gravity (buoyancy) waves in the atmosphere. The atmosphere contains a stack of large eddies. Vertical mixing of overlying environmental air into the large eddy volume occurs by turbulent eddy fluctuations [32 - 35, 37]. Townsend [42] has visualized the large eddy as the envelope enclosing turbulent eddies and the circulation speed of the large eddy is obtained as the integrated mean of the enclosed turbulent eddy fluctuations and is given as [32 - 35, 42].

$$W^2 = \frac{2}{\pi}\frac{r}{R}w_*^2 \qquad (1)$$

In the above Eq. (1) $W$ and $w_*$ are respectively the r.m.s (root mean square) circulation speeds of the large and turbulent eddies and $R$ and $r$ are their respective radii.

As seen from Figs. 1 and 2 and from the concept of eddy growth, vigorous counter flow (mixing) characterizes the large-eddy volume [32 - 35]. The total fractional volume dilution rate of the large eddy by vertical mixing across unit cross-section is derived from Eq. (1) [32 - 35, 37] and is given as follows.

$$k = \frac{w_*}{dW}\frac{r}{R} \qquad (2)$$

In Eq. (2) $w_*$ is the increase in vertical velocity per second of the turbulent eddy due to microscale fractional condensation (MFC) process and $dW$ is the corresponding increase in vertical velocity of large eddy.

The fractional volume dilution rate $k$ is equal to 0.4 for the scale ratio ($z$) $R/r =10$. Identifiable large eddies can exist in the atmosphere only for scale ratios more than 10 since, for smaller scale ratios the fractional volume dilution rate $k$ becomes more than half. Thus atmospheric eddies of various scales, i.e., convective, meso-, synoptic and planetary scale eddies are generated by successive decadic scale range eddy mixing process starting from the basic turbulence scale [32 - 35, 38].

From Eq.(2) the following logarithmic wind profile relationship for the *ABL* is obtained [32 - 35, 37].

$$W = \frac{w_*}{k} \ln z \tag{3}$$

The steady state fractional upward mass flux $f$ of surface air at any height $z$ can be derived using Eq. (3) and is given by the following expression [32 - 35, 37].

$$f = \sqrt{\frac{2}{\pi z}} \ln z \tag{4}$$

In Eq. (4) $f$ represents the steady state fractional volume of surface air at any level $z$. Since atmospheric aerosols originate from surface, the vertical profile of mass and number concentration of aerosols follow the $f$ distribution.

The model predicted aerosol vertical distributions are computed using Eq. (4) and are shown in Fig. 3. The model predicted profiles closely resemble the observed profiles reported by other investigators [23] and is computed as follows. Starting from primary eddy radius 100m near surface levels, successive length scale increments equal to 100m results in dominant eddy of radius 1000m after 10 growth steps. The next stage of dominant eddy growth starts with primary eddy length scale equal to 1000m. Ten successive length step growths (each equal to 1000m) results in the next stage large eddy of length scale equal to 10000m (10km) at the stratospheric levels. The semi-permanent peaks in the aerosol concentration at 1 km (*lifting condensation level*) and at about 10-15 km (*stratosphere*) identify the microscale capping inversion (MCI) at the crests of the convective and meso-scale eddies respectively. Earlier it was shown that the scale ratios for the convective and meso-scale eddies are respectively 10 and 100 with respect to the turbulence scale. Thus for the turbulent eddy of radius 100m, the MCI's for the convective and meso-scale eddies occur at 1 km and 10 km respectively.

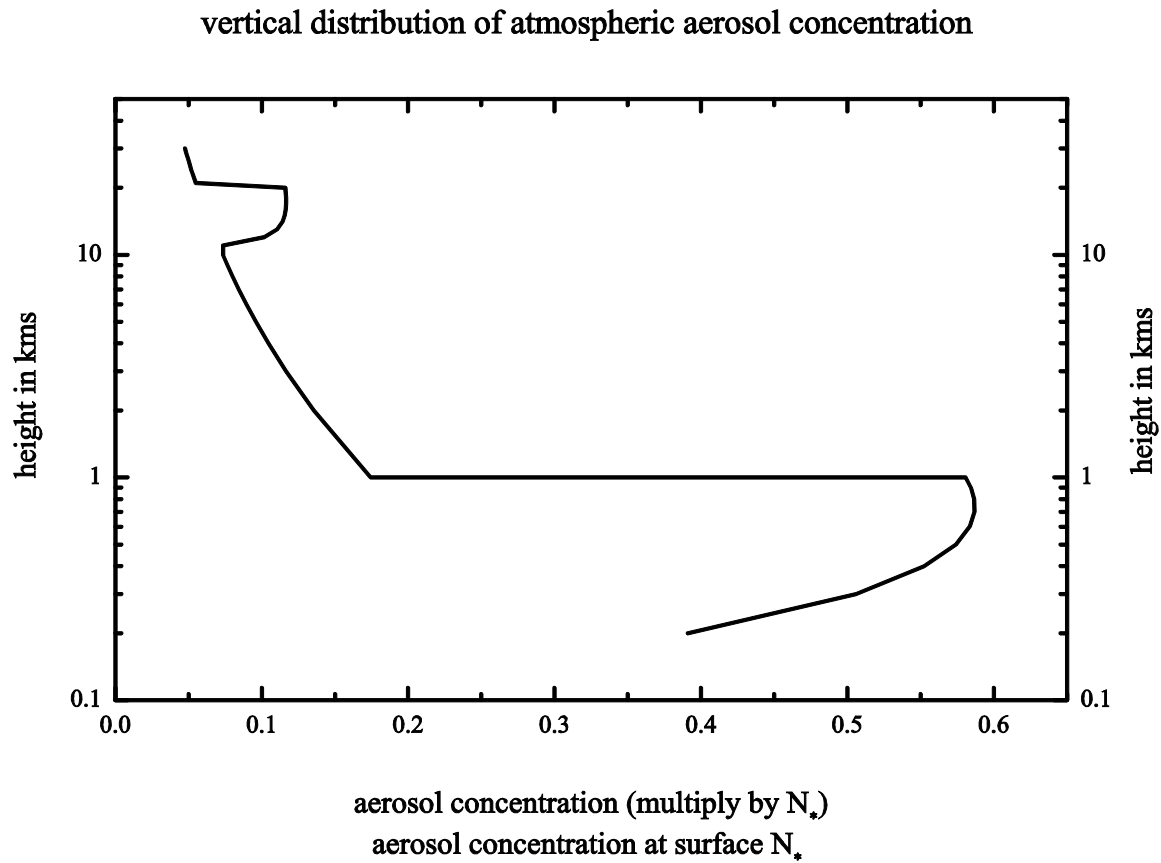

Fig. 3. Model predicted aerosol vertical distribution

The vertical mass exchange mechanism predicts the *f* distribution for the steady state vertical transport of aerosols at higher levels. Thus aerosol injection into the stratosphere by volcanic eruptions gives rise to the enhanced peaks in the regions of MCI in the stratosphere and other higher levels determined by the radius of the dominant turbulent eddy at that level.

### 4.1 Large eddy growth time

The time $\Gamma$ taken for the steady state aerosol concentration *f* to be established at the normalised height *z* is equal to the time taken for the large eddy to grow to the height *z* and is computed using the following relation.

$$\Gamma = \frac{r_*}{w_*}\sqrt{\frac{\pi}{2}}\,\text{li}\,\sqrt{z}$$

In the above equation *li* is the logarithm integral and the derivation of the above expression for $\Gamma$ is given in the following.

The time required for the large eddy of radius *R* to grow from the primary turbulence scale radius $r_*$ is computed as follows.

$$\text{The scale ratio } z = \frac{R}{r_*}$$

Therefore for constant turbulence radius $r_*$

$$dz = \frac{dR}{r_*}$$

The incremental growth d$R$ of large eddy radius is equal to

$$dR = r_* \, dz$$

The time d$t$ for the incremental cloud growth is expressed as follows

$$dt = \frac{dR}{W} = \frac{r_* \, dz}{W}$$

From the logarithmic wind profile relationship (Eq. 3) and the steady state fractional upward mass flux $f$ of surface air at any height $z$ (Eq. 4) the corresponding vertical velocity perturbation $W$ can be expressed in terms of the primary vertical velocity perturbation $w_*$ as [32]

$$W = w_* f z$$

$W$ may be expressed in terms of the scale ratio $z$ as given below

From Eq. 4
$$f = \sqrt{\frac{2}{\pi z}} \ln z$$

Therefore
$$W = w_* z \sqrt{\frac{2}{\pi z}} \ln z = w_* \sqrt{\frac{2z}{\pi}} \ln z$$

Therefore incremental eddy growth time d$t$ is given as

$$dt = \frac{r_* \, dz}{w_* f z} = \frac{r_* \, dz}{w_* z \sqrt{\frac{2}{\pi z}} \ln z}$$

The time $\Gamma$ taken for large eddy growth from surface to normalized height $z$ is obtained as

$$\Gamma = \int dt = \frac{r_*}{w_*} \sqrt{\frac{\pi}{2}} \int_2^z \frac{dz}{z^{1/2} \ln z}$$

The above equation can be written in terms of $\sqrt{z}$ as follows

$$d(z^{0.5}) = \frac{dz}{2\sqrt{z}}$$

$$dz = 2\sqrt{z} \, d(\sqrt{z})$$

Therefore

$$\Gamma = \frac{r_*}{w_*}\sqrt{\frac{\pi}{2}}\int_{x1}^{x2}\frac{d(\sqrt{z})}{\ln\sqrt{z}} = \frac{r_*}{w_*}\sqrt{\frac{\pi}{2}}\int_{x1}^{x2}\mathrm{li}(\sqrt{z}) \qquad (5)$$

$$x_1 = \sqrt{z_1} \text{ and } x_2 = \sqrt{z_2}$$

In the above equation $z_1$ and $z_2$ refer respectively to lower and upper limits of integration and $li$ is the Soldner's integral or the logarithm integral. The large eddy growth time $\Gamma$ can be computed from Eq. 5.

The vertical dispersion rate of aerosols/pollutants from known sources (e.g., volcanic eruptions, industrial emissions) can be computed using the relation for $f$ and $\Gamma$ (Eqs. 4 and 5).

## 5. Atmospheric aerosol size spectrum

### 5.1 Vertical variation of aerosol number concentration

The atmospheric eddies hold in suspension the aerosols and thus the size spectrum of the atmospheric aerosols is dependent on the vertical velocity spectrum of the atmospheric eddies as shown below.

From the logarithmic wind profile relationship (Eq. 3) and the steady state fractional upward mass flux $f$ of surface air at any height $z$ (Eq. 4) the vertical velocity $W$ can be expressed as [32]

$$W = w_* f z \qquad (6)$$

The corresponding moisture content $q$ at height $z$ is related to the moisture content $q_*$ at the surface and is given as (from Eq. 6)

$$q = q_* f z \qquad (7)$$

The aerosols are held in suspension by the eddy vertical velocity perturbations. Thus the suspended aerosol mass concentration $m$ at any level $z$ will be directly related to the vertical velocity perturbation $W$ at $z$, i.e., $W \sim mg$ where $g$ is the acceleration due to gravity. Therefore

$$m = m_* f z \qquad (8)$$

In Eq. (8) $m_*$ is the suspended aerosol mass concentration in the surface layer. Let $r_a$ and $N$ represent the mean volume radius and number concentration of aerosols at level $z$. The variables $r_{as}$ and $N_*$ relate to corresponding parameters at the surface levels. Substituting for the average mass concentration in terms of mean radius and number concentration

$$\frac{4}{3}\pi r_a^3 N = \frac{4}{3}\pi r_{as}^3 N_* f z \qquad (9)$$

The number concentration of aerosol decreases with height according to the $f$ distribution as shown earlier in Sec. 4 and Fig. 3 and is expressed as follows:

$$N = N_* f \qquad (10)$$

## 5.2 Vertical variation of aerosol mean volume radius

The mean volume radius of aerosol increases with height as shown in the following.

The velocity perturbation $W$ is represented by an eddy continuum of corresponding size (length) scales $z$. The aerosol mass flux across unit cross-section per unit time is obtained by normalizing the velocity perturbation $W$ with respect to the corresponding length scale $z$ to give the volume flux of air equal to $Wz$ and can be expressed as follows from Eq. (6):

$$Wz = (w_* fz)z = w_* fz^2 \qquad (11)$$

The corresponding normalised moisture flux perturbation is equal to $qz$ where $q$ is the moisture content per unit volume at level $z$. Substituting for $q$ from Eq. (7)

$$\text{normalised moisture flux at level } z = q_* fz^2 \qquad (12)$$

The moisture flux increases with height resulting in increase of mean volume radius of cloud condensation nuclei CCN because of condensation of water vapour. The corresponding CCN (aerosol) mean volume radius $r_a$ at height $z$ is given in terms of the aerosol number concentration $N$ at level $z$ and mean volume radius $r_{as}$ at the surface as follows from Eq. (12)

$$\frac{4}{3}\pi r_a^3 N = \frac{4}{3}\pi r_{as}^3 N_* fz^2 \qquad (13)$$

Substituting for $N$ from Eq. (10) in terms of $N_*$ and $f$

$$\begin{aligned} r_a^3 &= r_{as}^3 z^2 \\ r_a &= r_{as} z^{2/3} \end{aligned} \qquad (14)$$

The mean aerosol size increases with height according to the cube root of $z^2$ (Eq. 14). As the large eddy grows in the vertical, the aerosol size spectrum extends towards larger sizes while the total number concentration decreases with height according to the $f$ distribution. The atmospheric aerosol size spectrum is dependent on the eddy energy spectrum and may be expressed in terms of the recently identified universal characteristics of fractal fluctuations generic to atmospheric flows [35] as shown in Sec. 5.3 below.

## 5.3 Probability distribution of fractal fluctuations in atmospheric flows

The atmospheric eddies hold in suspension the aerosols and thus the size spectrum of the atmospheric aerosols is dependent on the vertical velocity spectrum of the atmospheric eddies. Atmospheric air flow is turbulent, i.e., consists of irregular fluctuations of all space-time scales characterized by a broadband spectrum of eddies. The suspended aerosols will also exhibit a broadband size spectrum closely related to the atmospheric eddy energy spectrum.

Atmospheric flows exhibit self-similar fractal fluctuations generic to dynamical systems in nature such as fluid flows, heart beat patterns, population dynamics, spread of forest fires, etc. Power spectra of fractal fluctuations exhibit inverse power law of form $f^{\alpha}$ where α is a constant indicating long-range space-time correlations or persistence. Inverse power law for power spectrum indicates scale invariance, i.e., the eddy energies at two different scales (space-time) are related to each other by a scale factor (α in this case) alone independent of the intrinsic properties such as physical, chemical, electrical etc of the dynamical system.

A general systems theory for turbulent fluid flows predicts that the eddy energy spectrum, i.e., the variance (square of eddy amplitude) spectrum is the same as the probability distribution $P$ of the eddy amplitudes, i.e. the vertical velocity $W$ values. Such a result that the additive amplitudes of eddies, when squared, represent the probabilities is exhibited by the subatomic dynamics of quantum systems such as the electron or photon. Therefore the unpredictable or irregular fractal space-time fluctuations generic to dynamical systems in nature, such as atmospheric flows is a signature of quantum-like chaos. The general systems theory for turbulent fluid flows predicts [32 - 35] that the atmospheric eddy energy spectrum follows inverse power law form incorporating the *golden mean* $\tau$ [35] and the normalized deviation $\sigma$ for values of $\sigma \geq 1$ and $\sigma \leq -1$ as given below

$$P = \tau^{-4\sigma} \qquad (15)$$

The vertical velocity $W$ spectrum will therefore be represented by the probability distribution $P$ for values of $\sigma \geq 1$ and $\sigma \leq -1$ given in Eq. (15) since fractal fluctuations exhibit quantum-like chaos as explained above.

$$W = P = \tau^{-4\sigma} \qquad (16)$$

Values of the normalized deviation $\sigma$ in the range $-1 < \sigma < 1$ refer to regions of primary eddy growth where the fractional volume dilution $k$ (Eq. 2) by eddy mixing process has to be taken into account for determining the probability distribution $P$ of fractal fluctuations (see Sec. 5.4 below).

### 5.4 Primary eddy growth region fractal space-time fluctuation probability distribution

Normalised deviation $\sigma$ ranging from $-1$ to $+1$ corresponds to the primary eddy growth region. In this region the probability $P$ is shown to be equal to $P = \tau^{-4k}$ (see below) where $k$ is the fractional volume dilution by eddy mixing (Eq. 2).

The normalized deviation $\sigma$ represents the length step growth number for growth stages more than one. The first stage of eddy growth is the primary eddy growth starting from unit length scale perturbation, the complete eddy forming at the tenth length scale growth, i.e., $R = 10r$ and scale ratio $z$ equals 10 [32]. The steady state fractional volume dilution $k$ of the growing primary eddy by internal smaller scale eddy mixing is given by Eq. (2) as

$$k = \frac{w_* r}{WR} \qquad (17)$$

The expression for $k$ in terms of the length scale ratio $z$ equal to $R/r$ is obtained from Eq. (1) as

$$k = \sqrt{\frac{\pi}{2z}} \qquad (18)$$

A fully formed large eddy length $R = 10r$ ($z=10$) represents the average or mean level zero and corresponds to a maximum of 50% probability of occurrence of either positive or negative fluctuation peak at normalized deviation σ value equal to zero by convention. For intermediate eddy growth stages, i.e., $z$ less than 10, the probability of occurrence of the primary eddy fluctuation does not follow conventional statistics, but is computed as follows taking into consideration the fractional volume dilution of the primary eddy by internal turbulent eddy fluctuations. Starting from unit length scale fluctuation, the large eddy formation is completed after 10 unit length step growths, i.e., a total of 11 length steps including the initial unit perturbation. At the second step ($z = 2$) of eddy growth the value of normalized deviation σ is equal to 1.1 - 0.2 (= 0.9) since the complete primary eddy length plus the first length step is equal to 1.1. The probability of occurrence of the primary eddy perturbation at this σ value however, is determined by the fractional volume dilution $k$ which quantifies the departure of the primary eddy from its undiluted average condition and therefore represents the normalized deviation σ. Therefore the probability density $P$ of fractal fluctuations of the primary eddy is given using the computed value of $k$ as shown in the following equation.

$$P = \tau^{-4k} \qquad (19)$$

The vertical velocity $W$ spectrum will therefore be represented by the probability density distribution $P$ for values of $-1 \leq \sigma \leq 1$ given in Eq. (19) since fractal fluctuations exhibit quantum-like chaos as explained above (Eq. 16).

$$W = P = \tau^{-4k} \qquad (20)$$

The probabilities of occurrence ($P$) of the primary eddy for a complete eddy cycle either in the positive or negative direction starting from the peak value (σ = 0) are given for progressive growth stages (σ values) in the following Table 1. The statistical normal probability density distribution corresponding to the normalized deviation σ values are also given in the Table 1.

Table 1: Primary eddy growth

| Growth step no | ± σ    | k     | Probability (%)  |                    |
|----------------|--------|-------|------------------|--------------------|
|                |        |       | Model predicted  | Statistical normal |
| 2              | .9000  | .8864 | 18.1555          | 18.4060            |
| 3              | .8000  | .7237 | 24.8304          | 21.1855            |
| 4              | .7000  | .6268 | 29.9254          | 24.1964            |
| 5              | .6000  | .5606 | 33.9904          | 27.4253            |
| 6              | .5000  | .5118 | 37.3412          | 30.8538            |
| 7              | .4000  | .4738 | 40.1720          | 34.4578            |
| 8              | .3000  | .4432 | 42.6093          | 38.2089            |
| 9              | .2000  | .4179 | 44.7397          | 42.0740            |
| 10             | .1000  | .3964 | 46.6250          | 46.0172            |
| 11             | 0      | .3780 | 48.3104          | 50.0000            |

The model predicted probability density distribution $P$ along with the corresponding statistical normal distribution with probability values plotted on linear and logarithmic scales respectively on the left and right hand sides are shown in Fig. 4. The model predicted probability distribution $P$ for fractal space-time fluctuations is very close to the statistical normal distribution for normalized deviation σ values less than 2 as seen on the left hand side of Fig. 4. The model predicts progressively higher values of probability $P$ for values of σ greater than 2 as seen on a logarithmic plot on the right hand side of Fig. 4.

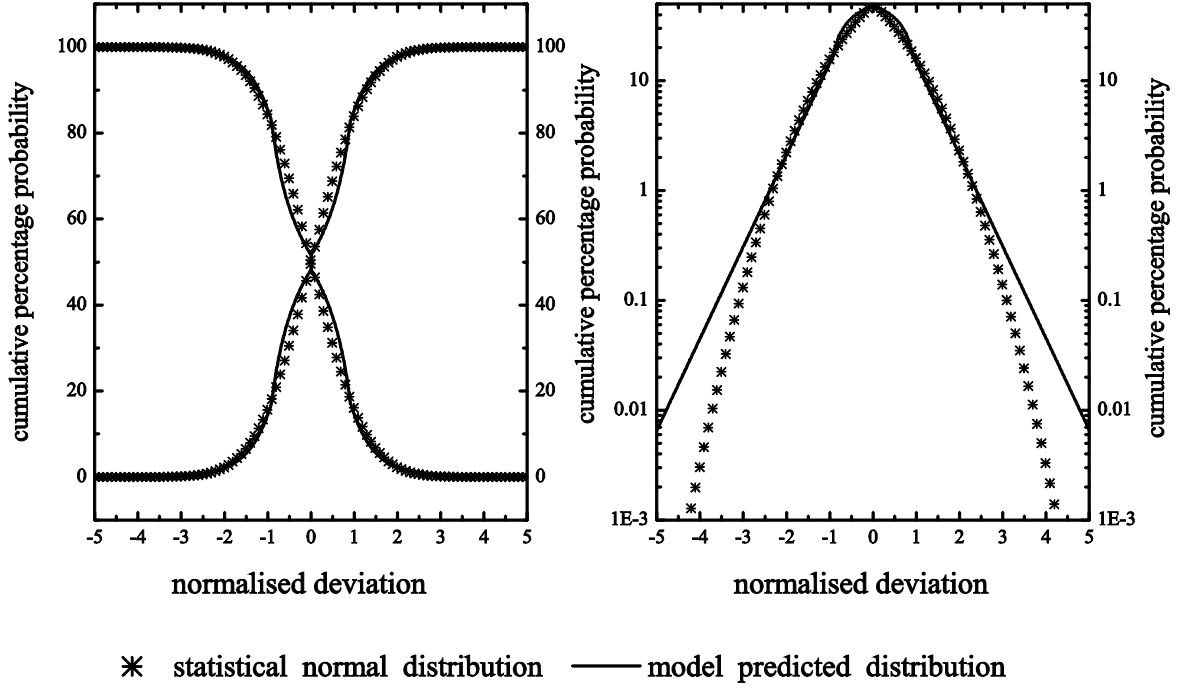

Fig. 4: Model predicted probability distribution *P* along with the corresponding statistical normal distribution with probability values plotted on linear and logarithmic scales respectively on the left and right hand sides.

### 5.5 Atmospheric wind spectrum and aerosol size spectrum

The steady state flux d$N$ of CCN at level $z$ in the normalised vertical velocity perturbation (d$W$)$z$ is given as

$$dN = N(dW)z \tag{21}$$

The logarithmic wind profile relationship for $W$ at Eq. (3) gives

$$dN = Nz\frac{w_*}{k}d(\ln z) \tag{22}$$

Substituting for $k$ from Eq. (2)

$$dN = Nz\frac{w_*}{w_*}Wz d(\ln z) = NWz^2 d(\ln z) \tag{23}$$

The length scale $z$ is related to the aerosol radius $r_a$ (Eq. 14). Therefore

$$\ln z = \frac{3}{2}\ln\left(\frac{r_a}{r_{as}}\right) \tag{24}$$

Defining a normalized radius $r_{an}$ equal to $\dfrac{r_a}{r_{as}}$, i.e., $r_{an}$ represents the CCN mean volume radius $r_a$ in terms of the CCN mean volume radius $r_{as}$ at the surface (or reference level). Therefore

$$\ln z = \frac{3}{2}\ln r_{an} \tag{25}$$

$$d\ln z = \frac{3}{2}d\ln r_{an} \tag{26}$$

Substituting for dln$z$ in Eq. (23)

$$dN = NWz^2\frac{3}{2}d(\ln r_{an}) \tag{27}$$

$$\frac{dN}{d(\ln r_{an})} = \frac{3}{2}NWz^2 \tag{28}$$

Substituting for $W$ from Eq. (16) and Eq. (20) in terms of the universal probability density $P$ for fractal fluctuations

$$\frac{dN}{d(\ln r_n)} = \frac{3}{2}NPz^2 \tag{29}$$

The general systems theory predicts that fractal fluctuations may be resolved into an overall logarithmic spiral trajectory with the quasiperiodic Penrose tiling pattern for the internal structure such that the successive eddy lengths follow the Fibonacci mathematical number series [32 - 35]. The eddy length scale ratio $z$ for length step $\sigma$ is therefore a function of the golden mean $\tau$ given as

$$z = \tau^\sigma \tag{30}$$

Expressing the scale length $z$ in terms of the golden mean $\tau$ in Eq. (29)

$$\frac{dN}{d(\ln r_{an})} = \frac{3}{2}NP\tau^{2\sigma} \tag{31}$$

In Eq. (31) $N$ is the steady state aerosol concentration at level $z$. The normalized aerosol concentration any level $z$ is given as

$$\frac{1}{N}\frac{dN}{d(\ln r_{an})} = \frac{3}{2}P\tau^{2\sigma} \tag{32}$$

The fractal fluctuations probability density is $P = \tau^{-4\sigma}$ (Eq. 16) for values of the normalized deviation $\sigma \geq 1$ and $\sigma \leq -1$ on either side of $\sigma = 0$ as explained earlier (Secs. 5.3, 5.4). Values of the normalized deviation $-1 \leq \sigma \leq 1$ refer to regions of primary eddy growth where the fractional volume dilution $k$ (Eq. 2) by eddy mixing process has to be taken into account for determining the probability density $P$ of fractal fluctuations. Therefore the probability density $P$ in the primary eddy growth region ($\sigma \geq 1$ and $\sigma \leq -1$) is given using the computed value of $k$ as $P = \tau^{-4k}$ (Eq. 20).

The normalised radius $r_{an}$ is given in terms of $\sigma$ and the golden mean $\tau$ from Eq. (25) and Eq. (30) as follows.

$$\ln z = \frac{3}{2} \ln r_{an}$$
$$r_{an} = z^{2/3} = \tau^{2\sigma/3}$$
(33)

The normalized aerosol size spectrum is obtained by plotting a graph of normalized aerosol concentration $\frac{1}{N}\frac{dN}{d(\ln r_{an})} = \frac{3}{2} P \tau^{2\sigma}$ (Eq. 32) versus the normalized aerosol radius $r_{an} = \tau^{2\sigma/3}$ (Eq. 33). The normalized aerosol size spectrum is derived directly from the universal probability density $P$ distribution characteristics of fractal fluctuations (Eq. 16 and Eq. 20) and is independent of the height $z$ of measurement and is universal for aerosols in turbulent atmospheric flows. The aerosol size spectrum is computed starting from the minimum size, the corresponding probability density $P$ (Eq. 32) refers to the cumulative probability density starting from 1 and is computed as equal to $P = 1 - \tau^{-4\sigma}$.

The universal normalised aerosol size spectrum represented by $\frac{1}{N}\frac{dN}{d(\ln r_{an})}$ versus $r_{an}$ is shown in Fig. 5.

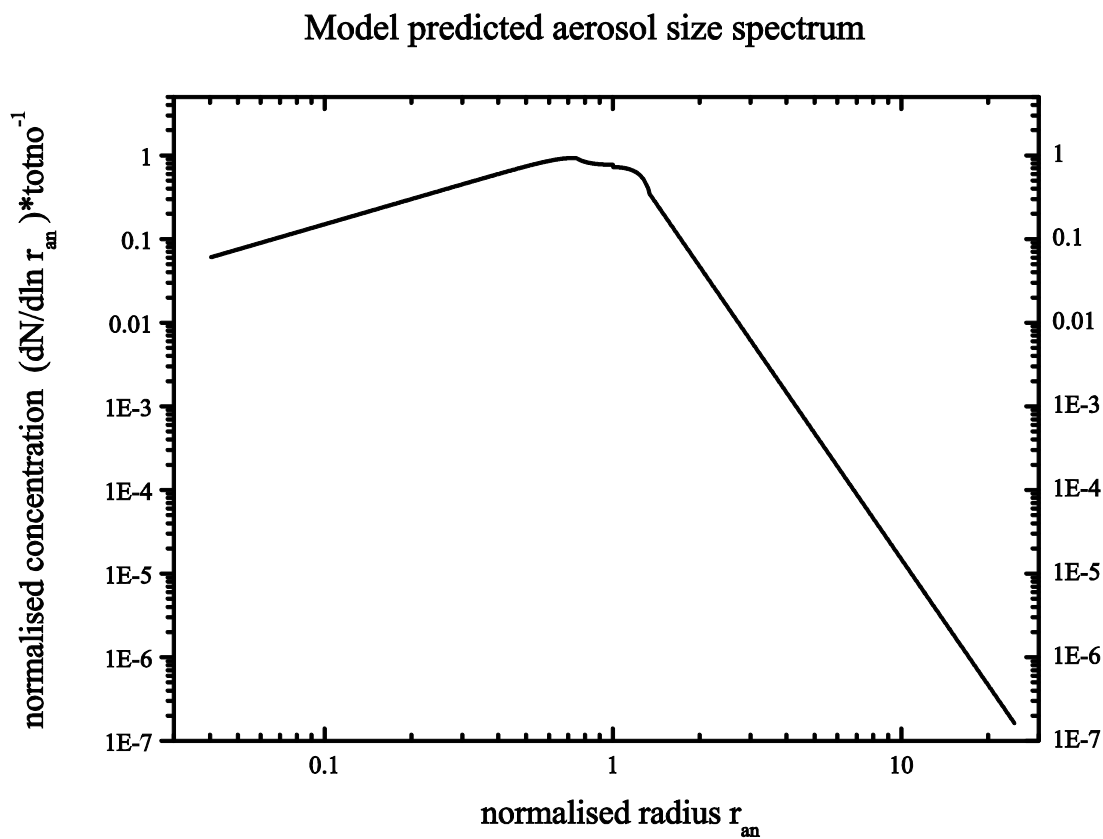

Fig. 5: Model predicted aerosol size spectrum

## 6. General systems theory and classical statistical physics

Lebowitz [26] has discussed the essential role of classical statistical mechanical concepts underlying the formulation of precise physical laws for observed macroscale phenomena in nature. Nature has a hierarchical structure, with time, length and energy scales ranging from the submicroscopic to the supergalactic. Surprisingly it is possible and in many cases essential to discuss these levels independently—quarks are irrelevant for understanding protein folding and atoms are a distraction when studying ocean currents. Nevertheless, it is a central lesson of science, very successful in the past three hundred years, that there are no new fundamental laws, only new phenomena, as one goes up the hierarchy. Thus, arrows of explanations between different levels always point from smaller to larger scales, although the origin of higher level phenomena in the more fundamental lower level laws is often very far from transparent. Statistical Mechanics (SM) provides a framework for describing how well-defined higher level patterns or behavior may result from the non-directed activity of a multitude of interacting lower level individual entities. The subject was developed for, and has had its greatest success so far in, relating mesoscopic and macroscopic thermal phenomena to the microscopic world of atoms and molecules. Statistical mechanics explains how macroscopic phenomena originate in the cooperative behavior of these microscopic particles [26].

The general systems theory visualizes the self-similar fractal fluctuations to result from a hierarchy of eddies, the larger scale being the space-time average of enclosed smaller scale eddies (Eq. 1) assuming constant values for the characteristic length scale $r$ and circulation speed $w_*$ throughout the large eddy space-time domain. The collective behavior of the ordered hierarchical eddy ensembles is manifested as the apparently irregular fractal fluctuations with long-range space-time correlations generic to dynamical systems. The concept that aggregate averaged eddy ensemble properties represent the eddy continuum belongs to 19$^{th}$ century classical statistical physics where the study of the properties of a system is reduced to a determination of average values of the physical quantities that characterize the state of the system as a whole [47] such as gases, e.g., the gaseous envelope of the earth, the atmosphere.

In classical statistical physics *kinetic theory of ideal gases* is a study of systems consisting of a great number of molecules, which are considered as bodies having a small size and mass [25]. Classical statistical methods of investigation [47, 25, 10, 30, 13, 14, 12, 6] are employed to estimate average values of quantities characterizing aggregate molecular motion such as mean velocity, mean energy etc., which determine the macro-scale characteristics of gases. The mean properties of ideal gases are calculated with the following assumptions. (1) The intra-molecular forces are completely absent instead of being small. (2) The dimensions of molecules are ignored, and considered as material points. (3) The above assumptions imply the molecules are completely free, move rectilinearly and uniformly as if no forces act on them. (4) The ceaseless chaotic movements of individual molecules obey Newton's laws of motion.

The Austrian physicist Ludwig Boltzmann suggested that knowing the probabilities for the particles to be in any of their various possible configurations would enable to work out the overall properties of the system. Going one step further, he also made a bold and insightful guess about these probabilities - that any of the many conceivable configurations for the particles would be equally probable. Boltzmann's idea works, and has enabled physicists to make mathematical models of thousands of real materials, from simple crystals to superconductors. It reflects the fact that many quantities in nature - such as the velocities of molecules in a gas - follow "normal" statistics. That is, they are closely grouped around the average value, with a "bell curve" distribution. Boltzmann's guess about equal probabilities only works for systems that have settled down to equilibrium, enjoying, for example, the same temperature throughout. The theory fails in any system where destabilizing external sources of energy are at work, such as the haphazard motion of turbulent fluids or the fluctuating energies of cosmic rays. These systems don't follow normal statistics, but another pattern instead [5].

Cohen [9] discusses Boltzmann's equation as follows. In 1872 when Boltzmann derived in his paper: *Further studies on thermal equilibrium between gas molecules* [4] what we now call the Boltzmann equation, he used, following Clausius and Maxwell, the assumption of 'molecular chaos', and he does not seem to have realized the statistical, i.e., probabilistic nature of this assumption, i.e., of the assumption of the independence of the velocities of two molecules which are going to collide. He used both a dynamical and a statistical method. However, Einstein strongly disagreed with Boltzmann's statistical method, arguing that a statistical description of a system should be based on the dynamics of the system. This opened the way, especially for complex systems, for other than Boltzmann statistics. It seems that perhaps a combination of dynamics and statistics is necessary to describe systems with complicated dynamics [9]. Sornette [40] discusses the ubiquity of observed power law distributions in complex systems as follows. The extension of Boltzmann's distribution to out-of-equilibrium systems is the subject of intense scrutiny. In the quest to characterize complex systems, two distributions have played a leading role: the normal (or Gaussian) distribution and the power law distribution. Power laws obey the symmetry of scale invariance. Power law distributions and more generally regularly varying distributions remain robust functional forms under a large number of operations, such as linear combinations, products, minima, maxima, order statistics, powers, which may also explain their ubiquity and attractiveness. Research on the origins of power law relations, and efforts to observe and validate them in the real world, is extremely active in many fields of modern science, including physics, geophysics, biology, medical sciences, computer science, linguistics, sociology, economics and more. Power law distributions incarnate the notion that extreme events are not exceptional. Instead, extreme events should be considered as rather frequent and part of the same organization as the other events [40].

In the following it is shown that the general systems theory concepts are equivalent to Boltzmann's postulates and the *Boltzmann distribution* with the inverse power law expressed as a function of the golden mean is the universal probability distribution function for the observed fractal fluctuations which corresponds closely to statistical normal distribution for moderate amplitude fluctuations and exhibit a fat long tail for hazardous extreme events in dynamical systems.

For any system large or small in thermal equilibrium at temperature $T$, the probability $P$ of being in a particular state at energy $E$ is proportional to $e^{-\frac{E}{K_B T}}$ where $K_B$ is the *Boltzmann's constant*. This is called the *Boltzmann distribution* for molecular energies and may be written as

$$P \propto e^{-\frac{E}{K_B T}} \qquad (34)$$

The basic assumption that the space-time average of a uniform distribution of primary small scale eddies results in the formation of large eddies is analogous to Boltzmann's concept of equal probabilities for the microscopic components of the system [5]. The physical concepts of the general systems theory (Sec. 4) enable to derive [36] *Boltzmann distribution* as shown in the following.

The r.m.s circulation speed *W* of the large eddy follows a logarithmic relationship with respect to the length scale ratio *z* equal to *R/r* (Eq. 3 ) as given below

$$W = \frac{w_*}{k} \log z$$

In the above equation the variable *k* represents for each step of eddy growth, the fractional volume dilution of large eddy by turbulent eddy fluctuations carried on the large eddy envelope [32] and is given as (Eq. 17)

$$k = \frac{w_* r}{WR}$$

Substituting for *k* in Eq. (17) we have

$$W = w_* \frac{WR}{w_* r} \log z = \frac{WR}{r} \log z$$

and                                                                                                                                                       (35)

$$\frac{r}{R} = \log z$$

The ratio *r/R* represents the fractional probability *P* of occurrence of small-scale fluctuations (*r*) in the large eddy (*R*) environment. Since the scale ratio *z* is equal to *R/r*, Eq. 35) may be written in terms of the probability *P* as follows.

$$\frac{r}{R} = \log z = \log\left(\frac{R}{r}\right) = \log\left(\frac{1}{(r/R)}\right)$$
$$P = \log\left(\frac{1}{P}\right) = -\log P$$
(36)

The maximum entropy principle concept of classical statistical physics is applied to determine the fidelity of the inverse power law probability distribution *P* (Eq. 16 and Eq. 20) for exact quantification of the observed space-time fractal fluctuations of dynamical systems ranging from the microscopic dynamics of quantum systems to macro-scale real world systems. Kaniadakis [24] states that the correctness of an analytic expression for a given power-law tailed distribution used to describe a statistical system is strongly related to the validity of the generating mechanism. In this sense the maximum entropy principle, the cornerstone of statistical physics, is a valid and powerful tool to explore new roots in searching for generalized statistical theories [24]. The concept of entropy is fundamental in the foundation of statistical physics. It first appeared in thermodynamics through the second law of thermodynamics. In statistical mechanics, we are interested in the disorder in the distribution of the system over the permissible microstates. The measure of disorder first provided by Boltzmann principle (known as Boltzmann entropy) is given by $S = K_B \ln M$, where $K_B$ is the thermodynamic unit of measurement of entropy and is known as Boltzmann constant. $K_B = 1.33 \times 10^{-16}$ erg/°C. *M*, called thermodynamic probability or statistical weight, is the total number of microscopic complexions compatible with the macroscopic state of the system and corresponds to the "degree of disorder" or 'missing information' [7]. For a probability distribution among a discrete set of states the generalized entropy for a system out of equilibrium is given as [7, 31, 2, 39].

$$S = -\sum_{j=1}^{\sigma} P_j \ln P_j \qquad (37)$$

In Eq. (37) $P_j$ is the probability for the $j^{th}$ stage of eddy growth in the present study, $\sigma$ is the length step growth which is equal to the normalized deviation and the entropy *S* represents the 'missing information' regarding the probabilities. Maximum entropy *S* signifies minimum preferred states associated with scale-free probabilities.

The validity of the probability distribution *P* (Eq. 16 and Eq. 20) is now checked by applying the concept of maximum entropy principle [24]. Substituting for log $P_j$ (Eq. 36) and for the probability $P_j$ in terms of the golden mean $\tau$ derived earlier (Eq. 16 and Eq. 20) the entropy *S* is expressed as

$$S = -\sum_{j=1}^{\sigma} P_j \log P_j = \sum_{j=1}^{\sigma} P_j^2 = \sum_{j=1}^{\sigma} \left( \tau^{-4\sigma} \right)^2$$

$$S = \sum_{j=1}^{\sigma} \tau^{-8\sigma} \approx 1 \text{ for large } \sigma \qquad (38)$$

In Eq. (38) *S* is equal to the square of the cumulative probability density distribution and it increases with increase in $\sigma$, i.e., the progressive growth of the eddy continuum and approaches 1 for large $\sigma$. According to the second law of thermodynamics, increase in entropy signifies approach of dynamic equilibrium conditions with scale-free characteristic of fractal fluctuations and hence the probability distribution *P* (Eq. 16 and Eq. 20) is the correct analytic expression quantifying the eddy growth processes visualized in the general systems theory.

In the following it is shown that the eddy continuum energy distribution *P* (Eq. 16 and Eq. 20) is the same as the *Boltzmann distribution* for molecular energies. From Eq. (35)

$$z = \frac{R}{r} = e^{\frac{r}{R}}$$

or $\qquad (39)$

$$\frac{r}{R} = e^{-\frac{r}{R}}$$

The ratio $r/R$ represents the fractional probability *P* (Eq. 16 and Eq. 20) of occurrence of small-scale fluctuations (*r*) in the large eddy (*R*) environment. Considering two large eddies of radii $R_1$ and $R_2$ ($R_2$ greater than $R_1$) and corresponding r.m.s circulation speeds $W_1$ and $W_2$ which grow from the same primary small-scale eddy of radius *r* and r.m.s circulation speed $w_*$ we have from Eq. (1)

$$\frac{R_1}{R_2} = \frac{W_2^2}{W_1^2}$$

From Eq. (39)

$$\frac{R_1}{R_2} = e^{-\frac{R_1}{R_2}} = e^{-\frac{W_2^2}{W_1^2}} \quad (40)$$

The square of r.m.s circulation speed $W^2$ represents eddy kinetic energy. Following classical physical concepts [25] the primary (small-scale) eddy energy may be written in terms of the eddy environment temperature $T$ and the *Boltzmann's constant* $K_B$ as

$$W_1^2 \propto K_B T \quad (41)$$

Representing the larger scale eddy energy as $E$

$$W_2^2 \propto E \quad (42)$$

The length scale ratio $R_1/R_2$ therefore represents fractional probability $P$ (Eq. 16 and Eq. 20) of occurrence of large eddy energy $E$ in the environment of the primary small-scale eddy energy $K_B T$ (Eq. 41). The expression for $P$ is obtained from Eq. (40) as

$$P \propto e^{-\frac{E}{K_B T}} \quad (43\ 37)$$

The above is the same as the *Boltzmann's equation* (Eq. 34).

The derivation of *Boltzmann's equation* from general systems theory concepts visualises the eddy energy distribution as follows: (1) The primary small-scale eddy represents the molecules whose eddy kinetic energy is equal to $K_B T$ as in classical physics. (2) The energy pumping from the primary small-scale eddy generates growth of progressive larger eddies [32]. The r.m.s circulation speeds $W$ of larger eddies are smaller than that of the primary small-scale eddy (Eq. 1). (3) The space-time *fractal* fluctuations of molecules (atoms) in an ideal gas may be visualized to result from an eddy continuum with the eddy energy $E$ per unit volume relative to primary molecular kinetic energy ($K_B T$) decreasing progressively with increase in eddy size.

The eddy energy probability distribution ($P$) of fractal space-time fluctuations also represents the *Boltzmann distribution* for each stage of hierarchical eddy growth and is given by Eq. 16 and Eq. 20 derived earlier, namely

$$P = \tau^{-4\sigma}$$

The general systems theory concepts are applicable to all space-time scales ranging from microscopic scale quantum systems to macroscale real world systems such as atmospheric flows.

## 7. Data

The following two data sets were used for comparison of observed with model predicted aerosol size spectrum.

### 7.1 Data set I

SAFARI 2000 CV-580 Aerosol Data, Dry Season 2000 (CARG). The Cloud and Aerosol Research Group (CARG) of the University of Washington participated in the SAFARI-2000 Dry Season Aircraft campaign with their Convair-580 research aircraft. This campaign covered five countries in southern Africa from August 10 through September 18, 2000. Various types of measurements were obtained on the thirty-one research flights of the Convair-580 to study their relationships to simultaneous measurements from satellites (particularly Terra), other research aircraft, and SAFARI-2000 ground-based measurements and activities. The [UW Technical Report for the SAFARI 2000 Project](http://daac.ornl.gov/data/safari2k/atmospheric/CV-580/comp/SAFARI-MASTER.pdf) gives a complete detailed guide to the extensive measurements obtained aboard the UW Convair-580 aircraft in support of SAFARI 2000 [17].

Aerosol Data (aer) Instrument details are given in Table 2

Table 2: Aerosol Data (aer) Instrument details

(ftp://ftp.daac.ornl.gov/data/safari2k/atmospheric/CV-580/comp/CV-580.pdf)

| | INSTRUMENT I |
|---|---|
| Name | pcaspdnc |
| Definition | Particle concentration between 0.1 and 3.0μm in 15 channels (corrected per micron) |
| Units | #/cc |
| Instrument | Particle Measuring Systems Model PCASP-100X |
| Processing | Calculated from raw counts and sample time |
| Notes | Channel Limits are: 0.10, 0.12, 0.14, 0.17, 0.20, 0.25, 0.30, 0.40, 0.50, 0.70, 0.90, 1.20, 1.50, 2.00, 2.50, 3.00 μm |
| | |
| | INSTRUMENT II |
| Name | tsidnc |
| Definition | Particle concentration between 0.5 and 20 μm in 52 channels (corrected per micron) |
| Units | #/cc |
| Instrument | TSI Model 3320 APS |
| Processing | Calculated from raw counts and sample time |
| Notes | There is no data from this instrument on UW flights 1815, 1818, 1820, and 1830. Channel Limits are: 0.487, 0.523, 0.562, 0.604, 0.649, 0.698, 0.750, 0.806, 0.866, 0.931, 1.000, 1.075, 1.155, 1.241, 1.334, 1.433, 1.540, 1.655, 1.778, 1.911, 2.054, 2.207, 2.371, 2.548, 2.738, 2.943, 3.162, 3.398, 3.652, 3.924, 4.217, 4.532, 4.870, 5.233, 5.623, 6.043, 6.494, 6.978, 7.499, 8.058, 8.660, 9.306, 10.000, 10.746, 11.548, 12.409, 13.335, 14.330, 15.399, 16.548, 17.783, 19.110, 20.535 μm |

Data sets satisfying the following four conditions were chosen for analysis. (i) Flights where both *pcasp* and *tsi3320* are available (ii) There are no unavailable (-999.99) data in any of the size intervals (iii) The number of class intervals with zero aerosol number concentration does not exceed three within the first five class intervals for *pcasp* for data inclusion for both *pcasp* and *tsi3320* (iv) The standard deviation of aerosol number concentrations in the class intervals of the size spectrum is less than 0.8 in order to exclude any abnormally large fluctuations, particularly in the tail end of the spectrum. Aerosol data sets for 21 aircraft flights were used for the study and details of available number of aerosol size spectra for each flight are given in Table 3. The available number of data values for each of the 15 size class intervals for *pcasp* and 52 size class intervals for *tsi3320* instrumentation systems are shown in Fig. 6.

Table 3: SAFARI 2000 CV-580 AEROSOL SIZE SPECTRA, DRY SEASON 2000 (CARG)

| \multicolumn{7}{c}{Details of available number of aerosol size spectra} |

| No | Flight | Year | Month | Day | pcasp - 15 class intervals | tsi3320 - 52 class intervals |
|---|---|---|---|---|---|---|
|  |  |  |  |  | No of spectra | No of spectra |
| 1 | 1810 | 2000 | 08 | 10 | 12192 | 10364 |
| 2 | 1812 | 2000 | 08 | 14 | 9200 | 8165 |
| 3 | 1813 | 2000 | 08 | 14 | 2615 | 1481 |
| 4 | 1814 | 2000 | 08 | 15 | 11615 | 4359 |
| 5 | 1816 | 2000 | 08 | 18 | 7738 | 6840 |
| 6 | 1819 | 2000 | 08 | 20 | 14801 | 13740 |
| 7 | 1821 | 2000 | 08 | 23 | 9865 | 6605 |
| 8 | 1823 | 2000 | 08 | 29 | 9016 | 7486 |
| 9 | 1824 | 2000 | 08 | 29 | 9329 | 9201 |
| 10 | 1825 | 2000 | 08 | 31 | 18709 | 13688 |
| 11 | 1826 | 2000 | 09 | 01 | 4409 | 4408 |
| 12 | 1828 | 2000 | 09 | 01 | 7004 | 5993 |
| 13 | 1829 | 2000 | 09 | 02 | 20153 | 19477 |
| 14 | 1831 | 2000 | 09 | 05 | 18806 | 18756 |
| 15 | 1832 | 2000 | 09 | 06 | 13275 | 12967 |
| 16 | 1833 | 2000 | 09 | 06 | 7449 | 7380 |
| 17 | 1834 | 2000 | 09 | 07 | 14376 | 4312 |
| 18 | 1835 | 2000 | 09 | 10 | 14727 | 13862 |
| 19 | 1836 | 2000 | 09 | 11 | 11988 | 9039 |
| 20 | 1838 | 2000 | 09 | 14 | 11529 | 3968 |
| 21 | 1839 | 2000 | 09 | 16 | 16241 | 7670 |
|  |  |  |  | Total | 245037 | 189761 |

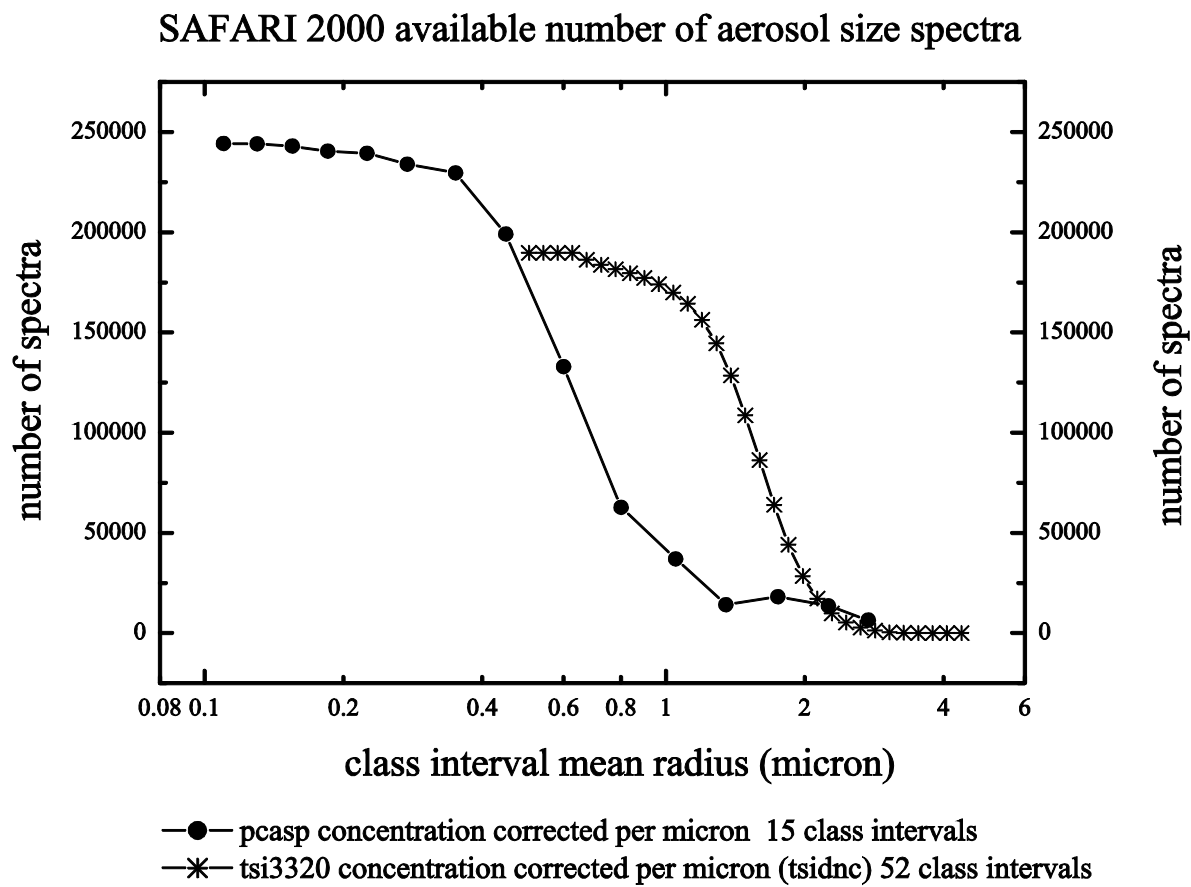

Fig 6. The available number of data values for each of the 15 size class intervals for *pcasp* and 52 size class intervals for *tsi3320* instrumentation systems.

### 7.2 Data Set II

Aerosol size distributions for three land stations (Ny Ålesund, Pallas and Hohenpeissenberg) were obtained from World Data Centre for Aerosols (http://wdca.jrc.it/data/parameters/data_size.html) at The Aerosol Size Distribution Data Archive and details of these data sets are given in Table 4. (Archive: ftp://ftp-ccu.jrc.it/pub/WDCA/NARSTO_archive/2.301/parameters/Size_Distribution). The annual means and corresponding standard deviations of aerosol number concentrations for the aerosol size class intervals were used for the study.

| \ | Table 4: World Data Centre Aerosols details |
|---|---|
| | Parameter: Number size distribution |
| | Station: Ny Ålesund |
| Data | QAC NARSTO PERMANENT ARCHIVE FILE NAME WDCA_GAWA_NYAZ_SIZEN_2001/2002/2003/2004.csv |
| Definition | Particle concentration between 0.017.8 and 0.7079 μm in 16 size classes |
| Units | #/cc |
| Method | Differential mobility particle sizer (DMA) |
| Start | 2000/01/01 |
| End | ongoing |
| Data period | 2001-2004 |
| Size class | 0.017.8, 0.022.4, 0.028.2, 0.0355, 0.0447, 0.0562, 0.0708, 0.0891, 0.1122, |
| limits | 0.1413, 0.1778, 0.2239, 0.2818, 0.3548, 0.4467, 0.5623, 0.7079 μm |
| | |
| Station: Pallas | |
| Data | QAC NARSTO PERMANENT ARCHIVE FILE NAME WDCA_GAWA_PAL_SIZEN_2001/2002/2003/2004.csv |
| Definition | Particle concentration between 0.0069 and 0.53 μm in 30 size classes |
| Units | #/cc |
| Method | Differential mobility particle sizer (DMA) |
| Start | 2000/01/01 |
| End | ongoing |
| Data period | 2001-2004 |
| Size class limits | 0.0069, 0.0079, 0.0092, 0.0106, 0.0123, 0.0142, 0.0164, 0.019, 0.022, 0.0254, 0.0294, 0.034, 0.0393, 0.0454, 0.0525, 0.0607, 0.0703, 0.0811, 0.0937, 0.1084, 0.1252, 0.1446, 0.1671, 0.1931, 0.223, 0.2573, 0.2974, 0.3436, 0.3966, 0.4583, 0.53 μm |
| | |
| Station: Hohenpeissenberg | |
| Data | QAC NARSTO PERMANENT ARCHIVE FILE NAME WDCA_GAWA_HOP_SIZE_2001/2002/2003/2004/2005.csv |
| Definition | Particle concentration between 0.1 and 6.75 μm in 15 size classes |
| Units | #/cc |
| Method | Optical particle counter (OPC) |
| Start | 1996/10/01 |
| End | ongoing |
| Data period | 2001-2005 |
| Size class limits | 0.1, 0.12, 0.15, 0.2, 0.25, 0.35, 0.45, 0.6, 0.75, 1, 1.5, 2, 3, 4.5, 6, 7.5μm |

## 8. Analysis

The aerosol size spectrum is given as (Sec. 5.5) the normalized aerosol number concentration equal to $\frac{1}{N}\frac{dN}{d(\ln r_{an})}$ versus the normalized aerosol radius $r_{an}$, where (i) $r_{an}$ is equal to $\frac{r_a}{r_{as}}$, $r_a$ being the mean class interval radius and $r_{as}$ the mean volume radius for the aerosol size spectrum 9reference level) (ii) $N$ is the total aerosol number concentration and $dN$ is the aerosol number concentration in the aerosol radius class interval $dr_a$ (iii) $d(\ln r_{an})$ is equal to $\frac{dr_a}{r_a}$ for the aerosol radius class interval $r_a$ to $r_a+dr_a$.

### 8.1 Data I: SAFARI 2000 CV-580 aerosol size spectra, dry season 2000 (CARG)

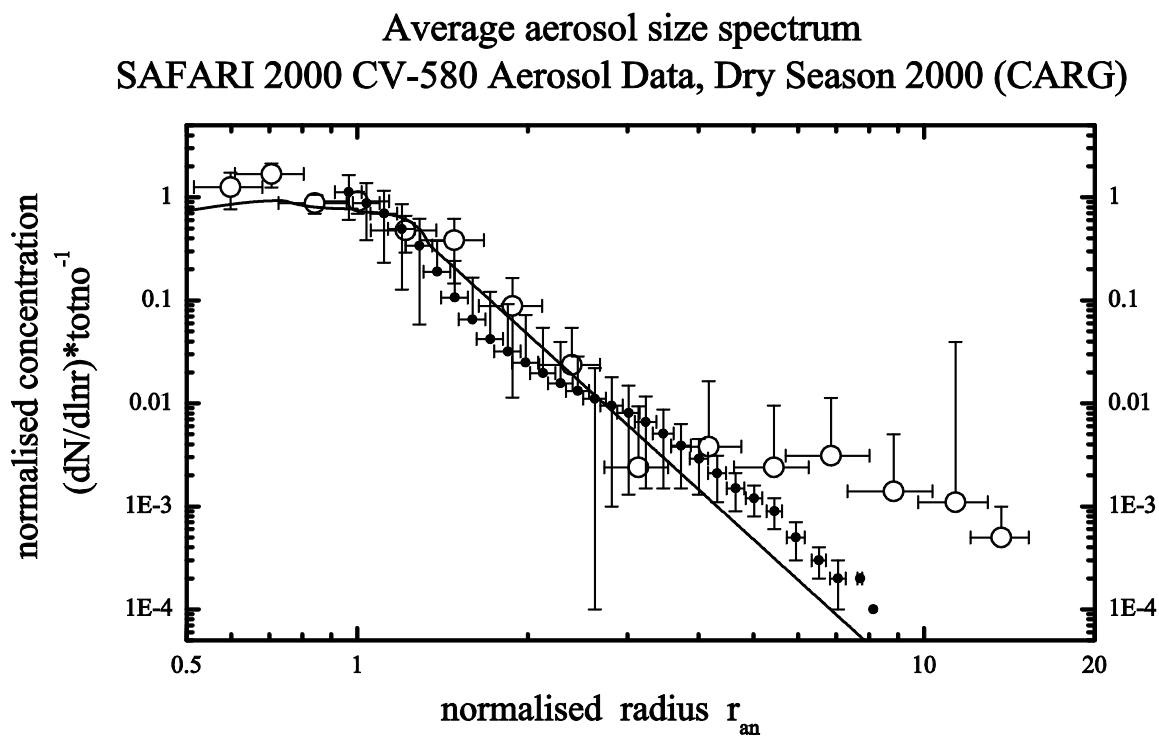

Fig. 7 Average aerosol size spectrum for SAFARI 2000 CV-580 aerosol size spectra and comparison with model prediction. Error bars indicate one standard deviation on either side of the mean

The mean and standard deviation of normalised aerosol size spectrum was computed for 245037 and 189761 individual spectra (Table 3) respectively for *pcasp* and *tsi3320* aerosol measurement instrument systems and shown in Fig. 7 along with the model predicted universal normalized aerosol size spectrum.

### 8.2 Data II: World Data Centre for Aerosols

The annual mean normalized aerosol size spectra with associated standard deviation were computed for the three stations (Table 4) Ny Ålesund, Pallas and Hohenpeissenberg for each year and shown in Fig. 8.

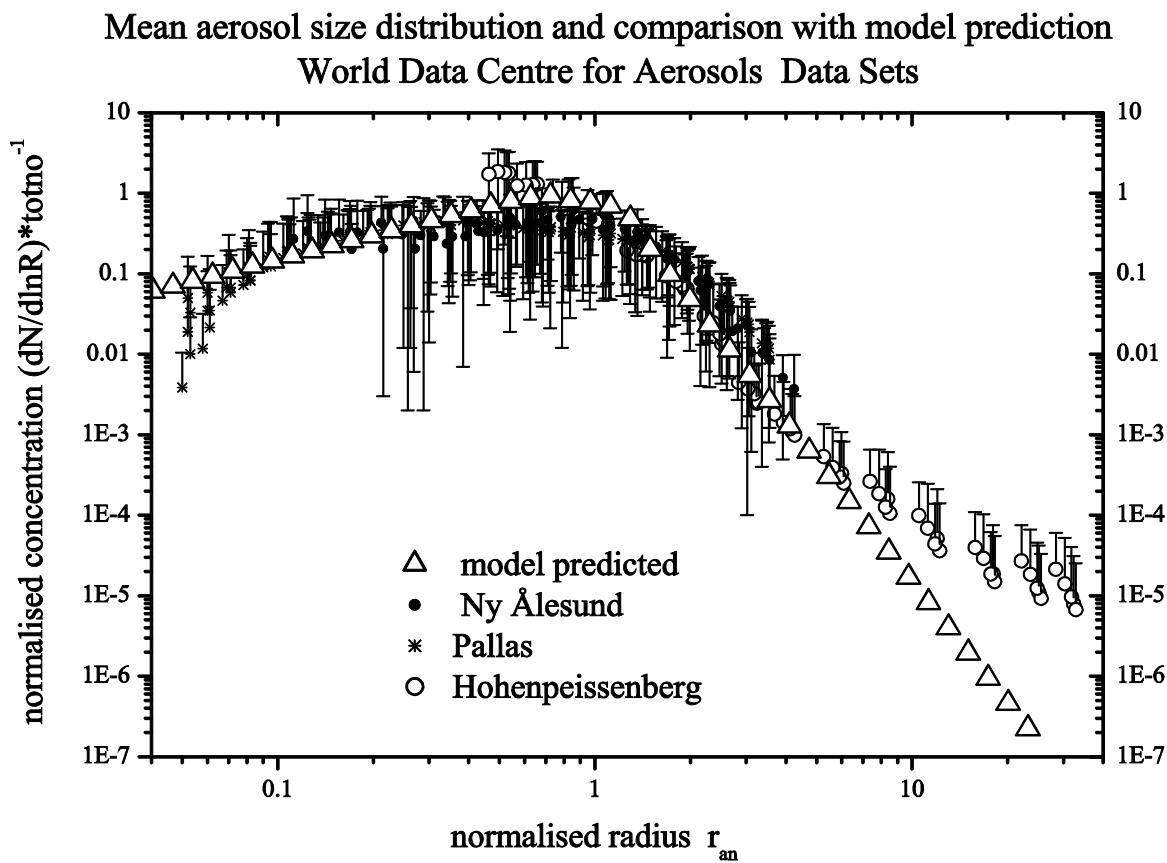

Fig. 8 Mean aerosol size spectrum for World data center for aerosols data sets and comparison with model prediction. Error bars indicate one standard deviation on either side of the mean

## 9. Conclusion

There is a close agreement between the model predicted and the observed aerosol size distributions for the two aerosol data sets (SAFARI 2000 and World Data Center) used in the study. SAFARI 2000 aerosol size distributions reported by Haywood *et al.* [15] also show similar shape for the distributions.

The distribution of atmospheric aerosols in not only determined by turbulence, but also by dry and wet chemistry, sedimentation, gas to particle conversion, coagulation, (fractal) variability at the surface, amongst others. However, at any instant the mass (and therefore the radius for homogeneous aerosols) size distribution of atmospheric suspensions (aerosols) is directly related to the wind velocity (eddy energy) spectrum which is shown to be universal (scale independent). The source for aerosols in the fine mode (less than 1 μm) and coarse mode (greater than 1 μm) are different and may account for the observed good fit of the observed radius size spectrum mostly for the fine aerosol mode only'

The general systems theory relating to the dynamics of the atmospheric eddy systems proposed in the present paper can be extended to other planetary, solar and stellar atmospheres.

## Acknowledgement

The author is grateful to Dr. A. S. R. Murty for encouragement during the course of the study.

GENERAL SYSTEMS THEORY FOR ATMOSPHERIC FLOWS					31

# APPENDIX I

# LIST OF FREQUENTLY USED SYMBOLS

| | |
|---|---|
| $d$ | aerosol diameter |
| $N$ | aerosol number concentration |
| $N_*$ | surface (or initial level) aerosol number concentration |
| $r_a$ | aerosol radius |
| $r_{as}$ | mean volume radius of aerosol size spectrum |
| $r_{an}$ | normalized radius equal to $r_a/r_{as}$ |
| $\alpha$ | exponent of inverse power law |
| $W$ | circulation speed (root mean square) of large eddy |
| $w$ | circulation speed (root mean square) of turbulent eddy |
| $R$ | radius of the large eddy |
| $r$ | radius of the turbulent eddy |
| $z$ | eddy length scale ratio equal to $R/r$; also represents normalized height |
| $w_*$ | primary (initial stage) turbulent eddy circulation speed |
| $r_*$ | primary (initial stage) turbulent eddy radius |
| $T$ | temperature $^{\circ}$K |
| $\Gamma$ | time period of large eddy circulation |
| $t$ | time period of turbulent eddy circulation |
| $k$ | fractional volume dilution rate of large eddy by turbulent eddy fluctuations |
| $f$ | steady state fractional upward mass flux of surface (or initial level) air |
| $q$ | moisture content at height $z$ |
| $q_*$ | moisture content at primary (initial stage) level |
| $m$ | suspended aerosol mass concentration at any level $z$ |
| $m_*$ | suspended aerosol mass concentration at primary (initial stage) level |
| $P$ | probability density distribution of fractal fluctuations |
| $\sigma$ | normalized deviation, large eddy length step growth |
| $K_B$ | Boltzmann's constant |
| $E$ | large eddy energy |